\documentclass[aps,showpacs,preprintnumbers,amsmath,amssymb,twocolumn]{revtex4}

\input{epsf}

\def\al{\alpha} \def\be{\beta} \def\ga{\gamma} \def\de{\delta}
   
\def\th{\theta}   \def\ka{\kappa}
\def\la{\lambda}   
\def\si{\sigma}   
   
\def\om{\omega}   
   
 \def\Om{\Omega} \def\mn{{\mu\nu}} 

\def\beq{\begin{equation}} \def\eeq{\end{equation}}
\def\beqa{\begin{eqnarray}} \def\eeqa{\end{eqnarray}}

\begin{document}

\title{Non-minimal curvature-matter couplings in modified gravity}

\author{Orfeu Bertolami}
\email{orfeu@cosmos.ist.utl.pt}
\affiliation{Instituto Superior T\'ecnico \\
             Departamento de F\'\i sica and Instituto de Plasmas e Fus\~ao Nuclear,\\
             Av. Rovisco Pais 1, 1049-001 Lisboa, Portugal}

\author{Tiberiu Harko}
\email{harko@hkucc.hku.hk} \affiliation{Department of Physics and
Center for Theoretical
             and Computational Physics, The University of Hong Kong,
             Pok Fu Lam Road, Hong Kong}

\author{Francisco S. N. Lobo}
\email{francisco.lobo@port.ac.uk} \affiliation{Institute of
Cosmology \& Gravitation,
             University of Portsmouth, Portsmouth PO1 2EG, UK}
\affiliation{Centro de Astronomia e Astrof\'{\i}sica da
Universidade de Lisboa, Campo Grande, Ed. C8 1749-016 Lisboa,
Portugal}

\author{Jorge P\'aramos}
\email{jorge.paramos@ist.utl.pt}
\affiliation{Instituto Superior T\'ecnico \\
             Departamento de F\'\i sica and Instituto de Plasmas e Fus\~ao Nuclear,\\
             Av. Rovisco Pais 1, 1049-001 Lisboa, Portugal}

\date{\today}

\begin{abstract}

Recently, in the context of $f(R)$ modified theories of gravity,
it was shown that a curvature-matter coupling induces a
non-vanishing covariant derivative of the energy-momentum,
implying non-geodesic motion and, under appropriate conditions, leading to the
appearance of an extra force. We study the implications of this
proposal and discuss some directions for future research.

\end{abstract}


\maketitle


\section{Introduction}

Current experimental evidence indicates that gravitational physics
is in agreement with Einstein's theory of General Relativity (GR)
to considerable accuracy (for thorough discussions see
\cite{Solar}); however, quite fundamental questions suggest that
it is unlikely that GR stands as the ultimate description of
gravity. Actually, difficulties arise from various corners, most
particularly in connection to the strong gravitational field
regime and the existence of spacetime singularities. Quantization
is a possible way to circumvent these problems, nevertheless,
despite the success of gauge field theories in describing the
electromagnetic, weak, and strong interactions, the description of
gravity at the quantum level is still missing, despite outstanding
progress achieved, for instance, in the context of
superstring/M-theory.

Furthermore, in fundamental theories that attempt to include
gravity, new long-range forces often arise in addition to the
Newtonian inverse-square law. Even if one assumes the validity of
the Equivalence Principle, Einstein's theory does not provide the
most general way to establish the spacetime metric. There are also
important reasons to consider additional fields, especially scalar
fields. Although the latter appear in unification theories, their
inclusion predicts a non-Einsteinian behaviour of gravitating
systems. These deviations from GR include
violations of the Equivalence Principle, modification of
large-scale gravitational phenomena, and variation of the
fundamental couplings.

On large scales, recent cosmological observations lead one to
conclude that our understanding of the origin and evolution of the
Universe based on GR requires that most of the energy content of
the Universe is in the form of currently unknown dark matter and
dark energy components that may permeate much, if not all
spacetime. Indeed, recent Cosmic Microwave Background Radiation
(CMBR) data indicate that our Universe is well described, within
the framework of GR, by a nearly flat Robertson-Walker metric.
Moreover, combination of CMBR, supernovae, baryon acoustic
oscillation and large scale structure data are consistent with
each other only if, in the cosmic budget of energy, dark energy
corresponds to about $73\%$ of the critical density, while dark
matter to about $23\%$ and baryonic matter to only about $4\%$.
Several models have been suggested to address issues related to
these new dark states. For dark energy, one usually considers the
so-called ``quintessence'' models, which involves the slow-roll
down of a scalar field along a smooth potential, thus inducing the
observed accelerated expansion (see \cite{Copeland:2006wr} for a
review). For dark matter, several weak-interacting particles
(WIMPs) have been suggested, many arising from extensions to the
Standard Model ({\it e.g.} axions, neutralinos). A scalar field
can also account for an unified model of dark energy and dark
matter \cite{Rosenfeld}. Alternatively, one can implement this
unification through an exotic equation of state, such as the
generalized Chaplygin gas \cite{Chaplygin}.

However, recently a different approach has attracted some
attention, namely the one where one considers a generalization of
the action functional. The most straight forward approach consists
in replacing the linear scalar curvature term in the
Einstein-Hilbert action by a function of the scalar curvature,
$f(R)$. In this context, a renaissance of $f(R)$ modified theories
of gravity has recently been verified in an attempt to explain the
late-time accelerated expansion of the Universe (see for instance
Refs. \cite{scalar,Sotiriou:2008rp} for recent
reviews). One could alternatively, resort to other scalar
invariants of the theory and necessarily analyze the observational
signatures and the parameterized post-Newtonian (PPN) metric
coefficients arising from these extensions of GR. In the context
of dark matter, the possibility that the galactic dynamics of
massive test particles may be understood without the need for dark
matter was also considered in the framework of $f(R)$ gravity
models \cite{darkmatter}.
Despite the extensive literature on these $f(R)$ models, an
interesting possibility has passed unnoticed till quite recently.
It includes not only a non-minimal scalar curvature term in the
Einstein-Hilbert Lagrangian density, but also a non-minimal
coupling between the scalar curvature and the matter Lagrangian
density \cite{Bertolami:2007gv} (see also Ref.
\cite{R-matt-coupling} for related discussions). It is interesting
to note that nonlinear couplings of matter with gravity were
analyzed in the context of the accelerated expansion of the
Universe \cite{Odintsov}, and in the study of the cosmological
constant problem \cite{Lambda}. In this contribution we discuss
various aspects of this proposal.

This work is organized as follows: in the following Section, the
main features of this novel model are presented. In Section III,
the issue of the degeneracy of Lagrangian densities, actually a
feature well known in GR
\cite{Schutz:1970my,Brown:1992kc,HawkingEllis}, is addressed in
the context of the new non-minimally coupled model
\cite{Bertolami:2008ab}. In Section IV, the scalar-tensor
representation of the model is presented, with particular emphasis
on the new features and difficulties encountered in the new model.
These issues are quite relevant, as they allow one to properly
obtain the PPN parameters $\beta$ and $\gamma$ and show that they
are consistent with the observations \cite{Bertolami:2008im}.
Section V, addresses the compatibility of the model with the
astrophysical condition for stellar equilibrium
\cite{Bertolami:2008vu}. In Section VI, a further generalization
of the model is discussed and an upper bound on the extra
acceleration introduced by the new non-minimal coupling is
obtained \cite{Ha08}. Finally, in Section VI our conclusions are
presented and objectives for further research are discussed.

Throughout this work, the convention $8\pi G=1$ and
the metric signature $(-,+,+,+)$ are used.

\section{Linear curvature-matter couplings}\label{Sec:II}

The action for curvature-matter couplings, in $f(R)$ modified
theories of gravity \cite{Bertolami:2007gv}, takes the following
form
\begin{equation} \label{model} S=\int \left[\frac{1}{2}f_1(R)+\left[1+\lambda
f_2(R)\right]{\cal L}_{m}\right] \sqrt{-g}\;d^{4}x\,,
\end{equation}
where $f_i(R)$ (with $i=1,2$) are arbitrary functions of the
curvature scalar $R$ and ${\cal L}_{m}$ is the Lagrangian density
corresponding to matter and $\la$ is a constant. Since the matter
Lagrangian is not modified in the total action, these may be
called modified gravity models with a non-minimal coupling between
matter and geometry.

Varying the action with respect to the metric $g_{\mu \nu }$
yields the field equations, given by
\begin{eqnarray}
&& F_1R_{\mu \nu }-\frac{1}{2}f_1g_{\mu \nu }-\nabla_\mu
\nabla_\nu F_1+g_{\mu\nu}\Box F_1= (1+\lambda f_2)T_{\mu\nu }
   \nonumber \\
&& -2\lambda F_2{\cal L}_m R_{\mu\nu}+2\lambda(\nabla_\mu
\nabla_\nu-g_{\mu\nu}\Box ){\cal L}_m F_2 \,, \label{field}
\end{eqnarray}
where one denotes $F_i(R)=f'_i(R)$, and the prime denotes
differentiation with respect to the scalar curvature. The matter
energy-momentum tensor is defined as
\begin{equation} T_{\mu \nu}=-\frac{2}{\sqrt{-g}}{\delta(\sqrt{-g}{\cal
L}_m)\over \delta(g^{\mu\nu})} \,. \label{defSET} \end{equation}
Now, taking into account the generalized Bianchi identities, one
deduces the following generalized covariant conservation equation
\begin{equation} \nabla^\mu T_{\mu \nu }={\lambda F_2\over 1+\lambda
f_2}\left[g_{\mu\nu}{\cal L}_m- T_{\mu \nu}\right]\nabla^\mu R\,.
\label{cons1} \end{equation}
It is clear that the non-minimal coupling between curvature and
matter yields a non-trivial exchange of energy and momentum between the
geometry and matter fields \cite{Bertolami:2008im}.

Considering, for instance, the energy-momentum tensor for a
perfect fluid,
\begin{equation} T_{\mu \nu }=\left( \rho +p\right) U_{\mu }U_{\nu }+pg_{\mu
\nu }\,, \end{equation}
where $\rho$ is the energy density and $p$ is the pressure,
respectively. The four-velocity, $U_{\mu }$, satisfies the
conditions $U_{\mu }U^{\mu }=-1$ and $U^{\mu }U_{\mu ;\nu }=0$.
Introducing the projection operator $h_{\mu \nu}=g_{\mu\nu}+U_{\mu
}U_{\nu}$, one can show that the motion is non-geodesic, and
governed by the following equation of motion for a fluid element
\begin{equation} {dU^{\mu}\over ds}+\Gamma _{\alpha
\beta}^{\mu}U^{\alpha}U^{\beta} =f^{\mu}\,, \label{eq1}
\end{equation}
where the extra force, $f^{\mu}$, appears and is given by
\begin{equation} f^{\mu}={1\over \rho +p}\left[{\lambda F_2\over 1+\lambda
f_2}\left({\cal L}_m-p\right)\nabla_\nu R+\nabla_\nu p \right]
h^{\mu\nu }\,.
     \label{force}
\end{equation}
One verifies that the first term vanishes for the specific choice
of ${\cal L}_m=p$, as noted in \cite{Sotiriou:2008it}.
However, as pointed out in \cite{Bertolami:2008ab}, this is
not the unique choice for the Lagrangian density of a perfect
fluid, as will be outlined below.

\section{Perfect fluid Lagrangian description}\label{Sec:III}

The novel coupling in action (\ref{model}) has attracted some
attention and, in a recent paper \cite{Sotiriou:2008it}, this
possibility has been applied to distinct matter contents.
Regarding the latter, it was argued that a ``natural choice'' for
the matter Lagrangian density for perfect fluids is ${\cal
L}_m=p$, based on \cite{Schutz:1970my,Brown:1992kc}, where
$p$ is the pressure. This specific choice implies the vanishing of
the extra force. However, although ${\cal L}_m=p$ does indeed
reproduce the perfect fluid equation of state, it is not unique:
other choices include, for instance, ${\cal L}_m=-\rho$
\cite{Brown:1992kc,HawkingEllis}, where $\rho$ is the energy
density, or ${\cal L}_m=-na$, where $n$ is the particle number
density, and $a$ is the physical free energy defined as
$a=\rho/n-Ts$, with $T$ being the fluid temperature and $s$ the
entropy per particle.

In this section, following \cite{Brown:1992kc,Bertolami:2008ab},
the Lagrangian formulation of a perfect fluid in the context of GR
is reviewed. The action is presented in terms of Lagrange
multipliers along the Lagrange coordinates $\alpha^A$ in order to
enforce specific constraints, and is given by
\begin{equation}
  S_m =\int d^4x \left[-\sqrt{-g} \;\rho(n,s)+J^\mu\left(\varphi_{,\mu}
  +s\theta_{,\mu}+\beta_A \alpha^{A}_{,\mu}\right)\right] ~.
  \label{fluid-action}
\end{equation}
Note that the action
$S_m=S(g_{\mu\nu},J^\mu,\varphi,\theta,s,\alpha^A,\beta_A)$ is a
functional of the spacetime metric $g_{\mu\nu}$, the entropy per
particle $s$, the Lagrangian coordinates $\alpha^A$, and spacetime
scalars denoted by $\varphi$, $\theta$, and $\beta_A$, where the
index $A$ takes the values 1, 2, 3 (see \cite{Brown:1992kc}
for details).

The vector density $J^\mu$ is interpreted as the flux vector of
the particle number density, and defined as
$J^\mu=\sqrt{-g}\,nU^\mu$. The particle number density is given by
$n=|J|/\sqrt{-g}$, so that the energy density is a function
$\rho=\rho(|J|/\sqrt{-g},s)$. The scalar field $\varphi$ is
interpreted as a potential for the chemical free energy $f$, and
is a Lagrange multiplier for $J^{\mu}_{,\mu}$, the particle number
conservation. The scalar fields $\beta_A$ are interpreted as the
Lagrange multipliers for $\alpha^{A}_{,\mu} J^\mu=0$, restricting
the fluid $4-$velocity to be directed along the flow lines of
constant $\alpha^{A}$.

The variation of the action with respect to $J^\mu$, $\varphi$,
$\theta$, $s$, $\alpha^A$ and $\beta_A$, provides the equations of
motion, which are not written here (we refer the reader to Ref.
\cite{Bertolami:2008ab} for details). Varying the action with
respect to the metric, and using the definition given by Eq.
(\ref{defSET}), provides the stress-energy tensor for a perfect
fluid
\begin{equation}
T^{\mu\nu}=\rho \, U^\mu U^\nu +\left(n{\partial \rho\over
\partial n}-\rho\right)\left(g^{\mu\nu}+U^\mu U^\nu \right)\,,
\end{equation}
with the pressure defined as
\begin{equation}
\label{pressure
definition} p=n{\partial \rho\over \partial n}-\rho\,.
\end{equation}
This definition of pressure is in agreement with the First Law of
Thermodynamics, $d\rho=\mu\, dn + nT ds$. The latter shows that
the equation of state can be specified by the energy density
$\rho(n,s)$, written as a function of the number density and
entropy per particle. The quantity $\mu=\partial \rho/\partial
n=(\rho+p)/n$ is defined as the chemical potential, which is the
energy gained by the system per particle injected into the fluid,
maintaining a constant sample volume and entropy per particle $s$.

Taking into account the equations of motions and the definitions
$J^\mu=\sqrt{-g}\,nU^\mu$ and $\mu=(\rho+p)/n$, the action Eq.
(\ref{fluid-action}) reduces to the on-shell Lagrangian density
${\cal L}_{m(1)} = p$, with the action given by \cite{Bertolami:2008ab}
\begin{equation}
  S_m =\int d^4x \sqrt{-g}\, p \,,
  \label{fluid-actionP}
\end{equation}
which is the form considered in Ref. \cite{Schutz:1970my}. It was
a Lagrangian density given by ${\cal L}_m=p$ that the authors of
\cite{Sotiriou:2008it} use to obtain a vanishing extra-force
due to the non-trivial coupling of matter to the scalar curvature
$R$. For concreteness, replacing ${\cal L}_m=p$ in Eq.
(\ref{force}), one arrives at the general relativistic expression
\begin{equation} f^{\mu}={h^{\mu\nu}\nabla_\nu p \over \rho+p}  ~.
\end{equation}

However, an on-shell degeneracy of the Lagrangian densities arises
from adding up surface integrals to the action. For instance,
consider the following surface integrals added to the action Eq.
(\ref{fluid-action}),
\begin{eqnarray} &-\int d^4x (\varphi J^\mu)_{,\mu} ~,~~~
-\int d^4x (\theta sJ^\mu)_{,\mu}\,,~~
     \nonumber   \\
& -\int d^4x (J^\mu \beta_A\alpha^A)_{,\mu}  ~,
  \nonumber
\end{eqnarray}
\noindent so that the resulting action takes the form
\begin{eqnarray}
  S&=&\int d^4x \Big[-\sqrt{-g}\, \rho(n,s)-\varphi J^{\mu}_{,\mu}
   \nonumber    \\
  &-&\theta(sJ^{\mu})_{,\mu}-\alpha^{A}(\beta_AJ^{\mu})_{,\mu}\Big] \,.
  \label{fluid-action2}
\end{eqnarray}
This action reproduces the equations of motion, and taking into
account the latter, the action reduces to \cite{Bertolami:2008ab}
\begin{equation}
  S_m =-\int d^4x \sqrt{-g}\, \rho \,,
  \label{fluid-action-rho}
\end{equation}
{\it i.e.}, the on-shell matter Lagrangian density takes the
following form ${\cal L}_m=-\rho$. This choice is also considered
for isentropic fluids, where the entropy per particle is constant
$s={\rm const.}$ \cite{Brown:1992kc,HawkingEllis}. For the latter,
the First Law of Thermodynamics indicates that isentropic fluids
are described by an equation of state of the form
$a(n,T)=\rho(n)/n-sT$ \cite{Brown:1992kc} (see Ref.
\cite{Bertolami:2008pa} for a bulk-brane discussion of this
choice).

For this specific choice of ${\cal L}_{m(2)}=-\rho$ the extra
force takes the following form:
\begin{equation}
f^{\mu}=\left(-{\lambda F_2\over 1+\lambda f_2}\nabla_\nu
R+{1\over \rho +p}\nabla_\nu p \right)h^{\mu\nu}\,.
    \label{force2}
\end{equation}
An interesting feature of Eq. (\ref{force2}) is that the term
related to the specific curvature-matter coupling is independent
of the energy-matter distribution.

The above discussion confirms that if one adopts a particular
on-shell Lagrangian density as a suitable functional for
describing a perfect fluid, then this leads to the issue of
distinguishing between different predictions for the extra force.
It is therefore clear that no straightforward conclusion may be
extracted regarding the additional force imposed by the
non-minimal coupling of curvature to matter, given the different
available choices for the Lagrangian density. One could even doubt
the validity of a conclusion that allows for different physical
predictions arising from these apparently equivalent Lagrangian
densities.

Despite the fact that the above Lagrangian densities ${\cal
L}_{m(i)}$ are indeed obtainable from the original action, it
turns out that they are not equivalent to the original Lagrangian
density ${\cal L}_{m}$. Indeed, this equivalence demands that not
only the equations of motion of the fields describing the perfect
fluid remain invariant, but also that the gravitational field
equations do not change. Indeed, the guiding principle behind the
proposal first put forward in Ref. \cite{Bertolami:2007gv} is to
allow for a non-minimal coupling between curvature and matter.

The modification of the perfect fluid action Eq.
(\ref{fluid-action}) should only affect the terms that show a
minimal coupling between curvature and matter, {\it i.e.}, those
multiplied by $\sqrt{-g}$ \cite{Bertolami:2008ab}. Thus, the
current density term, which is not coupled to curvature, should
not be altered. Writing ${\cal L}_c = -\rho(n,s)$, $V_\mu \equiv
\varphi_{,\mu}+s\theta_{,\mu}+\beta_A\alpha^{A}_{,\mu}$, for
simplicity, the modified action reads
\begin{equation}
  S'_m =\int d^4x \left[\sqrt{-g}\left[ 1 + \la f_2(R) \right]
  {\cal L}_c+J^\mu V_\mu + B^\mu_{;\mu} \right] \,,
\end{equation}
and one can see that only the non-minimal coupled term
${\cal L}_c$ appears in the field equations, as variations with
respect to $g^\mn$ of the remaining terms vanish:
\begin{eqnarray} \nonumber
&& F_1R_{\mu \nu }-{1\over 2}f_1g_{\mu \nu }-\nabla_\mu \nabla_\nu
F_1+g_{\mu\nu}\square F_1= (1+\lambda f_2)T_{\mu
\nu }\\ && -2\lambda F_2{\cal L}_c R_{\mu\nu}
   +2\lambda(\nabla_\mu
\nabla_\nu-g_{\mu\nu}\square){\cal L}_c F_2 \,. \label{field2}
\end{eqnarray}

Thus, quite logically, one finds that different
predictions for non-geodesic motion are due to different forms of
the gravitational field equations. Therefore, the equivalence
between different on-shell Lagrangian densities ${\cal L}_{m(i)}$
and the original quantity ${\cal L}_{m}$ is broken, so that one
can no longer freely choose between the available forms. For the
same reason, the additional extra force is unique, and obtained by
replacing ${\cal L}_c = -\rho$ into Eq. (\ref{force}), yielding
expression (\ref{force2}).

Indeed, in a recent paper \cite{Puetzfeld:2008xu}, a
generalization of the above approach is considered, by using a
systematic method that is not tied up to a specific choice of
matter Lagrangians. In particular, the propagation equations for
pole-dipole particles for a gravity theory with a very general
coupling between the curvature scalar and the matter fields is
examined, and it is shown that, in general, the extra-force does
not vanish.

\section{Scalar-tensor representation}\label{Sec:IV}

The connection between $f(R)$ theories of gravity and
scalar-tensor models with a ``physical'' metric coupled to the
scalar field is well known. In this section, one pursues the
equivalence between the model described by Eq. (\ref{model}) and
an adequate scalar-tensor theory. In close analogy with the
equivalence of standard $f(R)$ models \cite{analogy}, this
equivalence allows for the calculation of the PPN parameters
$\beta$ and $\gamma$ \cite{damour}.

One may first approach this equivalence by introducing two
auxiliary scalars $\psi$ and $\phi$ \cite{Sotiriou:2008it}, and
considering the following action
\begin{equation} \label{STmodel}
S_1=\int \left[\frac{1}{2}f_1(\phi)+\left[1+\lambda
f_2(\phi)\right]{\cal L}_{m}+\psi(R-\phi)\right]
\sqrt{-g}\;d^{4}x\,.
\end{equation}
Now, varying the action with respect to $\psi$ gives
$\phi = R$ and, consequently, action (\ref{model}) is recovered.
Varying the action with respect to $\phi$, yields
\begin{equation}\psi=\frac{1}{2}F_1+\lambda F_2{\cal L}_{m}\,.\end{equation}
Substituting this relationship back in (\ref{STmodel}), and
assuming that at least one of the functions $f_i$ is nonlinear in
$R$, one arrives at the following modified action
\begin{eqnarray} \label{STmodel2}
S_1&=&\int \Bigg[\frac{f_1(\phi)}{2}+\left[1+\lambda
f_2(\phi)\right]{\cal L}_{m}
    \nonumber   \\
&&+\left[\frac{1}{2}F_1(\phi)+\lambda F_2(\phi){\cal
L}_{m}\right](R-\phi)\Bigg] \sqrt{-g}\;d^{4}x\,,
\end{eqnarray}
where one still verifies the presence of the curvature-matter
coupling. Note that this is not an ordinary scalar-tensor theory,
due to the presence of the third and last terms. The former
represents a scalar-matter coupling, and the latter a novel
scalar-curvature-matter coupling. One may also use alternative
field definitions to cast the action (\ref{STmodel}) into a
Bran-Dicke theory with $\om=0$, {\it i.e.} no kinetic energy term
for the scalar field, but with the addition of a $R$-matter
coupling \cite{Sotiriou:2008it}. In conclusion, despite the fact
that the introduction of the scalar fields helps in avoiding the
presence of the nonlinear functions of $R$, the curvature-matter
couplings are still present and, consequently, these actions cannot
be cast into the form of a familiar scalar-tensor gravity
\cite{Sotiriou:2008it}.

However, one may instead pursue an equivalence with a theory with
not just one, but two scalar fields \cite{Bertolami:2008im}. This
is physically well motivated, since the non-minimal coupling of
matter and geometry embodied in Eq. (\ref{model}) gives rise to an
extra degree of freedom (notice that the case of a minimal
coupling $f_2 = 0$ yields $\psi = F_1(\phi) / 2$, so that this
degree of freedom is lost). Indeed, action of Eq. (\ref{STmodel})
may be rewritten as a Jordan-Brans-Dicke theory with a suitable
potential,
\begin{equation}\label{STmodel3}
S_1=\int \bigg[ \psi R - V(\phi,\psi) +\left[1+\lambda
f_2(\phi)\right]{\cal L}_{m}\bigg] \sqrt{-g}\;d^{4}x\,,
\end{equation}
with $V(\phi,\psi) = \phi \psi - f_1(\phi)/2$.

Variation of this action yields the field equations
\begin{eqnarray} && R_\mn - {1 \over 2} g_\mn R = 8 \pi G {1 + \la f_2(\phi) \over \psi} T_\mn \\ \nonumber &&  - {1 \over 2} g_\mn {V(\phi,\psi) \over \psi}  + {1 \over \psi} \left( \nabla_\mu \nabla_\nu - g_\mn \square \right) \psi  \end{eqnarray}
which, after the substitutions $\phi = R$ and $\psi = F_1/2+\lambda F_2{\cal L}_{m}$, collapses back to Eqs. (\ref{field}). Likewise, the Bianchi identities yield the generalized covariant conservation equation
\begin{eqnarray} && \nabla^\mu T_\mn =  {1 \over 1 + \la  f_2}\Bigg[ \left(\phi- R \right) \nabla_\nu \psi + \\ && \nonumber\left[ \left( \psi - \frac{1}{2} F_1 \right)g_\mn - \la F_2 T_\mn \right] \nabla^\mu \phi \Bigg] \,,\end{eqnarray}
also equivalent to Eq. (\ref{cons1}).


Through a conformal transformation $ g_\mn \rightarrow  g^*_\mn = \psi g_\mn$ (see {\it e.g.} \cite{conformal}), the scalar curvature can decouple from the scalar fields, so that the action is written in the so-called Einstein frame). A further redefinition of the scalar fields,
\begin{equation}\varphi^1 = \frac{\sqrt{3}}{2} \log \psi ~~~~,~~~~\varphi^2 = \phi\,,\end{equation}
allows the theory to be written canonically, that is,
\begin{eqnarray}
\label{actionfinal} S_1 &=& \int  \Bigg[  R^* -2 g^{*\mn} \si_{ij} \varphi^i_{,\mu}\varphi^j_{,\nu}  \\ \nonumber &&  -4 U(\varphi^1,\varphi^2) +\left[1 + \la f_2(\varphi^2)\right]{\cal L}_m^* \Bigg]   \sqrt{-g^*} ~d^4 x\,,
\end{eqnarray}
with ${\cal L}_m^* = {\cal L}_m / \psi^2 $, the redefined potential
\begin{eqnarray} U(\varphi^1, \varphi^2) &=&{1\over 4} \exp\left(-{2 \sqrt{3}\over 3}\varphi^1\right) \times \\ \nonumber && \left[ \varphi^2 - \frac{1}{2}f_1(\varphi^2) \exp\left(-{2 \sqrt{3}\over 3}\varphi^1\right) \right]\,, \end{eqnarray}
and the metric in the field space $(\varphi^1,\varphi^2)$,
\begin{equation}
\si_{ij} = \left(\begin{array}{cc}1 & 0 \\0 & 0\end{array}\right) ~~,
\end{equation}
which, after a suitable addition of an anti-symmetric part, will
be used to raise and lower Latin indexes.

Variation of action Eq. (\ref{actionfinal}) with respect to the
metric $g^*_\mn$ yields the field equations
\begin{eqnarray} \label{fieldfinal} && R^*_\mn - {1 \over 2} g^*_\mn R^* = 8 \pi G \left( 1 + \la f_2 \right) T^*_\mn + \\ \nonumber && \si_{ij} \left( 2\varphi^i_{,\mu}\varphi^j_{,\nu} - g^*_\mn g^{*\al\be} \varphi^i_{,\al}\varphi^j_{,\be} \right)- 2 g^*_\mn U  \,, \end{eqnarray}
while variation with respect to $\varphi^i$ gives the
Euler-Lagrange equations for each field:
\begin{equation}  \label{EL0} \square^* \varphi^i = B^i + 4 \pi G \left[ \al^i \left( 1+ \la f_2\right)  T^* - \la\si^{i2} F_2 {\cal L}^* \right]  \end{equation}
where one defines $B_i = \partial U / \partial \varphi^i $ and
\begin{equation} \label{alphas} \al_i = -\frac{1}{2} \frac{\partial \log \psi}{\partial \varphi^i } ~~\rightarrow~~ \al_1 = -\frac{\sqrt{3}}{3}~~~~,~~~~\al_2=0\,,\end{equation}

Eqs. (\ref{fieldfinal}), together with the Bianchi identities,
result in the generalized conservation law
\begin{equation} \label{noncons2} \nabla^{*\mu} T^*_\mn =  {\sqrt{3} \over 3} T^* \nabla^*_\nu \varphi^1 + {\la F_2 \over 1 + \la f_2} \left( g^*_\mn {\cal L}^* - T^*_{\mn}  \right) \nabla^{*\mu} \varphi^2\,. \end{equation}
%
%
From current bounds on the Equivalence Principle, it is reasonable to assume that the effect of the non-minimum
coupling of curvature to matter is weak, $\la f_2 \ll 1$.
Substituting this into (\ref{noncons2}) one gets, at zeroth-order
in $\la$,
\begin{equation} \nabla^{*\mu} T^*_\mn \simeq -\al_j T^* \varphi^j_{,\nu} \,, \label{zeroth} \end{equation}
so that one may disregard the $f_2(\varphi^2)$ factor in the
action (\ref{actionfinal}) and consider only through the coupling
present in $T^*$ (stemming from the definition of ${\cal L}_m^*$)
and the derivative of $\varphi^1$ (since $\varphi^1 \propto \log
\psi$ and $\psi = F_1 + F_2 {\cal L}$).

If both scalar fields are light, leading to long range
interactions, one may calculate the PPN parameters $\be$ and $\ga$
\cite{damour}, given by
\begin{equation} \be - 1 = {1 \over 2} \left[ {\al^i \al^j \al_{j,i} \over \left( 1 + \al^2 \right)^2 } \right]_0~~~~,~~~~ \ga - 1 = -2 \left[ {\al^2 \over 1+ \al^2 } \right]_0 \,, \label{PPNparameters} \end{equation}
\noindent where $\al_{j,i} = \partial \al_j / \partial \varphi^i$ and $\al^2  = \al_i \al^i = \si^{ij} \al_i \al_j $; the subscript $_0$ refers to the asymptotic value of the related quantities, which is connected to the cosmological values of the curvature and matter Lagrangian density. From the values found in Eq. (\ref{alphas}), one concludes that $\be = \gamma = 1$, as obtained in GR. However, it should be expected that small deviations of order $O(\la)$ arise when one considers the full impact of Eq. (\ref{noncons2}).

Furthermore, it should be empathized that the added degree of
freedom embodied in the non-minimal $ f_2 \neq 0$ coupling is
paramount in obtaining values for the PPN parameters $\be$ and
$\ga$ within the current experimental bounds (or, conversely,
allowing for future constraints of the magnitude of $\la$ and the
form of $f_2$); indeed, in the case where only the curvature term
is non-trivial, $f_1 \neq R $ and $f_2 = 0$, one degree of freedom
is lost and the parameter $\al \neq 0$ defined in Eq.
(\ref{alphas}) is no longer a vector, but a scalar quantity: as a
result, $\al^2 \neq 0$ and one gets $\ga = 1/2$. In the discussed
model, the vector $\al_i$ has $\al^2 = 0$, thus solving this
pathology (see \cite{Bertolami:2008im} and references therein
for a thorough discussion).

Finally, notice that these results should be independent of the
particular scheme chosen for the equivalence between the original
model and a scalar-tensor theory; this may be clearly seen by
opting for a more ``natural'' choice for the two scalar fields (in
the Jordan frame), such that $\phi= R$ and $\psi = {\cal L}$.
Although more physically motivated, this choice of fields is less
pedagogical and mathematically more taxing
\cite{Bertolami:2008im}.

\section{Implications for stellar equilibrium}\label{Sec:V}

In this section, one studies the impact of the non-minimally
coupled gravity model embodied in action Eq. (\ref{model}) in what
may be viewed as its natural proving ground: regions where
curvature effects may be high enough, to evidence some deviation
from GR, although moderate enough so these are still
perturbative -- a star \cite{Bertolami:2008vu} (see also
\cite{poly} for other physical examples of the adopted
methodology). As will be shown, the purpose of this exercise is
to calculate deviations to the central temperature of the Sun
(known with an accuracy of 6\%), due to the perturbative effect of
the non-minimal coupling of geometry to matter.

Clearly, a full treatment of the equations of motion (\ref{field})
is exceedingly demanding, unless a specific form for $f_1(R)$ and
$f_2(R)$ is considered. Furthermore, since one is mainly
interested in the ascertaining the effects of the non-minimal
coupling within a high curvature and pressure medium, the
modifications due to the pure curvature term $f_1$ should be
overwhelmed by the effect of  $f_2$; under such circumstances, one
may discard the former term, as thus take the trivial $f_1 = R$
case. A thorough discussion on the validity of this approximation
with regard to representative, physically viable candidates for
the function $f_1(R)$ is found in Ref. \cite{Bertolami:2008vu}.

One now deals with the particular form of the coupling function
$f_2$. One considers the simplest form, which might arise from the
first order expansion of a more general function in the weak field
environ of the Sun, $f_2 =R $ (this implies that $[\la] =M^{-2}$).
Also, one assumes that stellar matter is described by an ideal
fluid characterized by a Lagrangian density $\mathcal{L}_m = p$,
\cite{Schutz:1970my,Brown:1992kc}. Adopting $f_1 = f_2 = R $, the
field equations become
\beqa \label{fieldRR}&& \left( 1 + 2\la p \right) R_\mn - {1 \over
2}R \left( g_\mn +2 \la T_\mn \right) = \\* \nonumber && 2\la
(\nabla_\mu \nabla_\nu - g_\mn \square)p + {1 \over 2 }T_\mn ~~, \eeqa
Notice that both $\la p$ and $\la\rho$ are dimensionless
quantities: the perturbative condition $\la f_2 \ll 1$ translates
to $\la p \ll1$ and $\la \rho \ll 1$.

Taking the trace of the above equation yields
\beq R = {3p - \rho + 6 \la \square p \over 2 \left[1+\la(\rho - 5
p ) \right]} ~~, \label{ricci} \eeq inserting $T = T_\mu^\mu =
\rho - 3p$. Substituting this into Eq. (\ref{fieldRR}) and keeping
only first order terms in $\la$, one obtains
 \beqa \label{motion} && 2 [1 + \la (\rho - 3 p )] R_\mn = \\*
 \nonumber && (3p - \rho) g_\mn +2 (1-2\la p) T_\mn + 2 \la
 (4 \nabla_\mu \nabla_\nu - g_\mn \square) p ~~, \eeqa


Since temporal variations are assumed to occur at the cosmological
scale $H_0^{-1}$, and are thus negligible at an astrophysical time
scale, one considers an ideal, spherically symmetric system, with
a line element derived from the Birkhoff metric (in its
anisotropic form)
\beq ds^2 = e^{\nu(r)} dt^2 - \left( e^{\si(r)}dr^2 +d \Om^2
\right)~~, \label{line} \eeq with $d \Om = r^2 (d \th^2 + sin^2
\th ~d \phi^2)$. Following the usual treatment, one defines the
effective mass $m_e$ through $e^{-\si} = 1 - 2 Gm_e/ r$ which,
replacing in Eq. (\ref{motion}) yields, to first order in $\la$,
\beqa \label{mass'} && m'_e \approx 4 \pi r^2 \rho \left[ 1 + 2\la
\left(p - {\rho \over 2} - {3 \over 2} {p^2 \over \rho} \right)
\right] + \\* \nonumber && {\la r^2 \over 4 G} \left( 5 e^{-\nu}
\nabla_0 \nabla_0 + 3e^{-\si} \nabla_r \nabla_r + 2{ \nabla_\th
\nabla_\th \over
r^2} \right) p ~~, \eeqa which clearly shows the perturbation to
the gravitational mass, defined by $m'_g = 4 \pi r^2 \rho$ (in
here, the prime denotes differentiation with respect to $r$).

Taking the Newtonian limit \beq r \gg 2Gm_e(r)~,~ \rho(r) \gg
p(r)~,~ m_e(r) \gg 4 \pi p(r) r^3 ~~, \label{newtonGR} \eeq and
going through a few algebraic steps (depicted in
\cite{Bertolami:2008vu}), one eventually obtains the
non-relativistic hydrostatic equilibrium equation
\beq \label{hydroeq} p' + {Gm_e \rho\over r^2 } = 2\la \left[
\left( \left[ {5 \over 8} p'' - 4 \pi G p \rho \right] r - {p'
\over 4} \right) \rho+ p \rho' \right] ~~. \eeq where the
perturbation introduced by the non-minimal coupling is clearly
visible.


In order to scrutinize the profile of pressure and density inside
the Sun, one requires a suitable equation of state. Instead of
pursuing a realistic representation of the various layers of the
solar structure, one resorts to a very simplistic assumption, the
so-called polytropic equation of state. This is commonly given by
$p = K \rho^{(n+1)/n}$, where $K$ is the polytropic constant,
$\rho$ is the mass density and $n$ is the polytropic index. A
polytropic equation of state with $n = 3$ was used by Eddington in
his first solar model, and will be adopted here due.

Given this equation of state, one may write $\rho = \rho_c \th^n
(\xi)$ and $p = p_c \th^{n+1} ( \xi) $, with $\xi = r / r_0$ a
dimensionless variable and $r_0^2 \equiv {(n+1)} p_c / 4 \pi
G\rho_c^2 $; $ \rho_c = 1.622 \times 10^5~{\rm kg/m}^3$ is the central
density, and $p_c = 2.48 \times 10^{16}~{\rm Pa}$ is the central
pressure. One obtains the perturbed Lane-Emden equation for the
function $\th(\xi)$:
\beqa \label{LE} && {1\over \xi^2} \Bigg[ \xi^2 \th' \bigg( 1 +
A_c \th^n \times \\ \nonumber && \bigg[ \left[{5 \over 8} \left(
\th'' + n {\th'^2 \over \th} \right) - N_c \th^{n+1} \right] { \xi
\over \th' } +   {3n-1 \over 4(n+1) } \bigg]\bigg) \Bigg]' = \\
\nonumber && - \th^n \left[1 + A_c \left( {3 \over 8}\left[ \th''
+ n {\th'^2 \over \th} \right] + { \th' \over 4 \xi} - {\th^{n }
\over 2} \right) \right] ~~,\eeqa where the prime now denotes
derivation with respect to the dimensionless radial coordinate
$\xi$, and one defines the dimensionless parameters $A_c \equiv
\la \rho_c$ and $N_c \equiv p_c / \rho_c = 1.7 \times 10^{-6}$,
for convenience. Clearly, setting $A_c = 0$ one recovers the
unperturbed Lane-Emden equation \cite{books}.

Notice that the perturbed Lane-Emden equation is a third-degree
differential equation; its numerical resolution is computationally
intensive and displays some complex behaviour; conveniently, the
assumed perturbative regime prompts for the expansion of the
function $\th(\xi) = \th_0 (1+A_c \de)$ around the unperturbed
solution $\th_0(\xi)$. Inserting this into Eq. (\ref{LE}) and
expanding to first-order in $A_c$, one obtains
\beqa\label{difeq} && \de'' + 2 \left( {\th_0' \over \th_0} + {1
\over \xi} \right) \de' + (n-1) \th_0^{n-1} \de = {5n\over 2} \xi
\th_0^{2 n-2} \th_0' \\* \nonumber && + (2 n +1) N_c \xi \th_0^{2
n-1} \th_0' + {9 n+5 \over 4 (n+1)} \th_0^{2 n-1} + 3 N_c
\th_0^{2n} \\* \nonumber && - {5n (n-1) \over 8} \xi \th_0^{n-3}
\th_0'^3+ {n(3n+7) \over 4(n+1)} \th_0^{n-2} \th_0'^2 + {1 \over
2}{\th_0^{n-1} \th_0' \over \xi} ~~.\eeqa supplemented by the
initial conditions $\de(0)=\de'(0)=0$. Notice that the choice for
the perturbative expansion leads to a solution $\de$ independent
from the parameter $A_c$.

\begin{figure}
\epsfysize=5.7cm
\epsffile{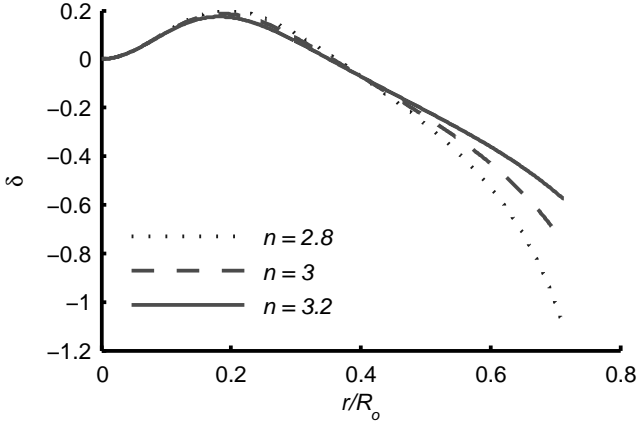}
\caption{Relative perturbation $\de$ for $ 2.8 \leq n \leq 3.2 $. }
\label{relprofiles}
\end{figure}

After dealing with the issue of exterior matching conditions and
bypassing a troublesome divergence of $\de$ near the boundary of
the star \cite{Bertolami:2008vu}, one may obtain the numerical
solution for Eq. (\ref{difeq}) for a polytropic index in the
vicinity of $n = 3$, as depicted in Fig.~\ref{relprofiles}.

\begin{figure}
\epsfysize=5.7cm \epsffile{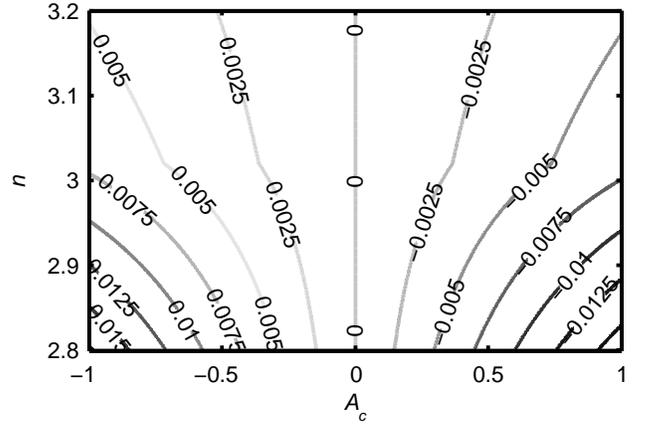} \caption{Relative
deviation of the central temperature $T_c / T_{c0}-1$, with
contour lines of step $0.1 \% $.} \label{relTc2d}
\end{figure}

Finally, one turns to the issue of calculating one of the
observables under scrutiny, that is, the central temperature of
the Sun. The polytropic equation of state indicates that $ \rho
\propto T^{n+1} $, which yields
\beqa && 1 - \left({T_{c0} \over T_c}\right)^{n+1} = \\* \nonumber
&& {A_c \over \xi_r^2 \th_{0r}' } \int_0^{\xi_r} \xi^2 \th_0^n
\left[n \de + {3 n \over 8} {\th_0'^2 \over \th_0} - {\th_0' \over
2 \xi} - {7 \over 8} \th_0^n \right] ~d\xi ~~. \eeqa where $\xi_r
= R_r / r_0$ and $R_r = 0.713 R_\odot$ marks the onset of the
convection zone (where the chosen equation of state fails) and
$T_{c0}$ is the central temperature derived from the $A_c=0$
unperturbed scenario. One may derive a parameter plot in the
$(n,A_c)$ parameter space, shown in Fig.~\ref{relTc2d}. As can be
seen, no relative deviation of the central temperature occurs
above the experimentally determined level of $6\%$. However, since
the values found are of the order of $1\%$, one may hope that any
future refinement of the experimental error of $T_c$ could yield a
direct bound on the parameter $A_c$. Furthermore, the perturbative
condition $\la \ll \ka \rho_c$ is confirmed (reintroducing the
factor $\ka$, for clarity), which translates to $|\la| \ll 4.24
\times 10^{33}~{\rm eV}^{-2} $.

\section{Models with arbitrary couplings between matter and geometry}
\label{Sec:VI}
The discussed gravity models with linear coupling between matter
and geometry, given by Eq.~(\ref{model}), can be further
generalized by assuming that the supplementary coupling between
matter and geometry takes place via an arbitrary function of the
matter Lagrangian ${\cal L}_{m}$, so that the action is given by
\cite{Ha08}
\begin{equation}\label{actgen}
S=\int \left\{ \frac{1}{2}f_{1}(R)+{\cal G}\left( {\cal L}_{m}\right) \left[1+\lambda f_{2}\left( R\right) \right]%
\right\} \sqrt{-g}d^{4}x,
\end{equation}
where ${\cal G}\left( {\cal L}_{m}\right) $ is an arbitrary
function of the matter Lagrangian density ${\cal L}_{m}$. The  action given by Eq.~(\ref{actgen})
represents the most general extension of the Einstein-Hilbert action for GR, $S=\int \left[ R/2+{\cal
L}_{m}\right] \sqrt{-g}d^{4}x$. For $f_{1}(R)=R$, $f_{2}(R)=0$ and
${\cal G}\left( {\cal L}_{m}\right) ={\cal L}_{m}$, one recovers GR. With $f_{2}(R)=0$ and $ {\cal
G}\left({\cal L}_{m}\right) ={\cal L}_{m}$ one obtains the $f(R)$
generalized gravity models. The case ${\cal G}\left( {\cal
L}_{m}\right) ={\cal L}_{m}$ corresponds to the linear coupling
between matter and geometry, given by Eq.~(\ref{model}). The only
requirement for $f_{i}$, $ i=1,2$ and ${\cal G}$ is that they are
analytical functions of the Ricci scalar $R$ and $ {\cal L}_{m}$,
respectively -- that is, they can be expressed as a Taylor series
expansion about any point.

The field equations corresponding to action (\ref{actgen}) are
\begin{eqnarray}
&&F_{1}(R)R_{\mu \nu }-\frac{1}{2}f_{1}(R)g_{\mu \nu }+\left( g_{\mu \nu
}\square -\nabla _{\mu }\nabla _{\nu }\right) F_{1}(R)=  \nonumber
\label{feq} \\
&&-2\lambda {\cal G}\left( \mathcal{L}_{m}\right) F_{2}(R)R_{\mu \nu }  \nonumber \\
&&-2\lambda \left( g_{\mu \nu }\square -\nabla _{\mu }\nabla _{\nu }\right)
{\cal G}\left( \mathcal{L}_{m}\right) F_{2}(R) \nonumber\\
&&-\left[ 1+\lambda f_{2}(R)\right] \left[ K\left( \mathcal{L}_{m}\right)
\mathcal{L}_{m}-{\cal G}\left( \mathcal{L}_{m}\right) \right] g_{\mu \nu }
\nonumber \\
&&-\left[ 1+\lambda f_{2}(R)\right] K\left( \mathcal{L}_{m}\right) T_{\mu
\nu },
\end{eqnarray}
where $F_{i}(R)=df_{i}(R)/dR$, $i=1,2$, and $K\left( \mathcal{L}_{m}\right)
=d{\cal G}\left( \mathcal{L}_{m}\right) /d\mathcal{L}_{m}$, respectively.

By taking the covariant divergence of Eq. (\ref{feq}), with the
use of the mathematical identity $\nabla ^{\mu }\left[ a^{\prime
}(R)R_{\mu \nu }-a(R)g_{\mu \nu }/2+\left( g_{\mu \nu }\square
-\nabla _{\mu }\nabla _{\nu }\right) a(R)\right] = 0$
\cite{Ko06}, where $a(R)$ is an arbitrary function of the Ricci
scalar and $a^{\prime }(R)=da/dR$, we obtain
\begin{eqnarray}
\nabla ^{\mu }T_{\mu \nu } &=&\nabla ^{\mu }\ln \left\{ \left[ 1+\lambda f_{2}(R)\right] K\left(
\mathcal{L}_{m}\right) \right\} \left( \mathcal{L}_{m}g_{\mu \nu }-T_{\mu
\nu }\right)   \nonumber \\
&=&2\nabla ^{\mu }\ln \left\{ \left[ 1+\lambda f_{2}(R)\right] K\left(
\mathcal{L}_{m}\right) \right\} \frac{\partial \mathcal{L}_{m}}{\partial
g^{\mu \nu }}.  \label{noncons}
\end{eqnarray}

For ${\cal G}\left( {\cal L}_{m}\right) ={\cal L}_{m}$, one
recovers the equation of motion of massive test particles in the
linear theory, Eq.~(\ref{force}). As a specific model of
generalized gravity models with arbitrary matter-geometry
coupling, one considers the case in which the matter Lagrangian
density is an arbitrary function of the energy density of the
matter $\rho $ only, so that ${\cal L}_{m}={\cal L}_{m}\left( \rho
\right) $. One assumes that during the hydrodynamic evolution the
energy density current is conserved, $\nabla _{\nu }\left( \rho
U^{\nu }\right) =0$. Then, the energy-momentum tensor of matter is
given by
\begin{equation}\label{tens}
T^{\mu \nu }=\rho \frac{d{\cal L}_m}{d\rho }U^{\mu }U^{\nu }+\left(
{\cal L}_m-\rho \frac{d{\cal L}_m}{d\rho }\right) g^{\mu \nu },
\end{equation}
where we have used the relation $\delta \rho =\left( 1/2\right)
\rho \left( g_{\mu \nu }-U_{\mu }U_{\nu }\right) \delta g^{\mu \nu
}$, a direct consequence of the conservation of the energy density
current.

The energy-momentum tensor given by Eq.~(\ref{tens}) can be
written in a form similar to the perfect fluid case if one assumes
that the thermodynamic pressure $p$ obeys a barotropic equation of
state, $p=p\left( \rho \right) $. In this case the matter
Lagrangian density and the energy-momentum tensor can be written
as
\begin{equation}
{\cal L}_{m}\left( \rho \right) =\rho \left[1+\Pi \left(\rho \right)\right]=\rho \left(1+\int_{0}^{p}\frac{dp}{\rho }\right)-p\left( \rho \right),
\end{equation}
and
\begin{equation}\label{tens1}
T^{\mu \nu }=\left\{ \rho \left[1+\Pi \left( \rho \right)\right]+p\left( \rho \right)  %
\right\} U^{\mu }U^{\nu }+p\left( \rho \right) g^{\mu \nu },
\end{equation}
respectively, where
 \begin{equation}
\Pi \left( \rho \right)
=\int_{0}^{p}\frac{dp}{\rho }-\frac{p\left(\rho \right)}{\rho }.
\end{equation}

Physically, $\Pi \left(\rho \right)$ can be interpreted as the
elastic (deformation) potential energy of the body, and therefore
Eq.~(\ref{tens1}) corresponds to the energy-momentum tensor of a
compressible elastic isotropic system. From Eq.~(\ref {noncons}),
one obtains the equation of motion of a test particle in the
modified gravity model with the matter Lagrangian an arbitrary
function of the energy density of matter as Eq.~(\ref{eq1}), where
the extra force is now given by
\begin{equation}
f^{\mu }=\nabla _{\nu }\ln \left\{ \left[ 1+\lambda f_{2}(R)\right]K\left[ {\cal L}_m\left( \rho \right) %
\right] \frac{d{\cal L}_{m}\left( \rho \right) }{d\rho }\right\} h^{\mu \nu }.
\end{equation}
It is easy to see that the extra-force $f^{\mu }$, generated due
to the presence of the coupling between matter and geometry, is
perpendicular to the four-velocity, $f^{\mu }U_{\mu }=0$. The
equation of motion, Eq.~ (\ref{eq1}), can be obtained from the
variational principle
\begin{equation}
\delta S_{p}=\delta \int {\cal L}_pds=\delta \int \sqrt{Q}\sqrt{g_{\mu
\nu }U^{\mu }U^{\nu }}ds=0,  \label{actpart}
\end{equation}
where $S_{p}$ and ${\cal L}_p=\sqrt{Q}\sqrt{g_{\mu \nu }U^{\mu }U^{\nu
}}$ are the action and Lagrangian density, respectively, and
\begin{equation}
\sqrt{Q}=\left[ 1+\lambda f_{2}(R)\right] K\left[ \mathcal{L}_{m}\left( \rho
\right) \right] \frac{d\mathcal{L}_{m}\left( \rho \right) }{d\rho }.
\label{Q}
\end{equation}

The variational principle Eq.~(\ref{actpart}) can be used to study
the Newtonian limit of the model. In the weak gravitational field
limit, $ds\approx \sqrt{1+2\phi -\vec{v}^{2}}dt\approx \left(
1+\phi -\vec{v}^{2}/2\right) dt$, where $\phi $ is the Newtonian
potential and $\vec{v}$ is the usual tridimensional velocity of
the particle. By representing the function $\sqrt{ Q}$ as
\begin{eqnarray}
\sqrt{Q}&=&\left[ 1+\lambda f_{2}(R)\right] K\left[ {\cal L}_{m}\left( \rho \right) \right]
\frac{d{\cal L}_{m}\left(
\rho \right) }{d\rho }\nonumber\\
&=&1+\Phi \left( R,{\cal L}_{m}\left( \rho \right) ,\frac{%
d{\cal L}_{m}\left( \rho \right) }{d\rho }\right) ,
\end{eqnarray}
where $\left|\Phi \right|<<1$, the equation of motion of a test
particle can be obtained from the variational principle
\begin{equation}
\delta \int \left[ \Phi \left( R,{\cal L}_{m}\left( \rho \right)
,\frac{d{\cal L}_{m}\left( \rho \right) }{d\rho }\right) +\phi
-\frac{\vec{v}^{2}}{2}\right] dt=0,
\end{equation}
and is given by
\begin{equation}
\vec{a}=-\nabla \phi -\nabla \Phi=\vec{a}_{N}+\vec{a}_{E},
\end{equation}
where $\vec{a}_{N}=-\nabla \phi $ is the usual Newtonian
gravitational acceleration and $\vec{a}_{E}=-\nabla \Phi$ a
supplementary effect induced by the coupling between
matter and geometry.

An estimative of the effect of the extra-force generated by the
coupling between matter and geometry on the orbital parameters of
planetary motion around the Sun can be obtained by using the
properties of the Runge-Lenz vector, defined as $\vec{A}=\vec{v} \times \vec{L}-\alpha \vec{e}_{r}$, where $\vec{v}$ is the
velocity relative to the Sun, with mass $M_{\odot}$,  of a planet
of mass $m$, $\vec{r}=r\vec{e}_{r}$ is the two-body position
vector, $\vec{p}=\mu \vec{v}$ is the relative momentum,  $\mu
=mM_{\odot }/\left( m+M_{\odot}\right) $ is the reduced mass, $\ \
\vec{L}=\vec{r} \times \vec{p}=\mu r^{2}\dot{\theta}\vec{k}$ is
the angular momentum, and $\alpha =GmM_{\odot}$ \cite{prec}. For
an elliptical orbit of eccentricity $e$, major semi-axis $a$, and
period $T$, the equation of the orbit is given by $\left(
L^{2}/\mu \alpha \right) r^{-1}=1+e\cos \theta $.  The Runge-Lenz
vector and its derivative can be expressed as
\begin{equation}
\vec{A}=\left( \frac{\vec{L}^{2}}{\mu r}-\alpha \right) \vec{e}_{r}-%
\dot{r}L\vec{e}_{\theta },
\end{equation}
and
\begin{equation}
\frac{d\vec{A}}{d\theta }=r^{2}\left[ \frac{dV(r)}{dr}-\frac{\alpha }
{r^{2}}\right] \vec{e}_{\theta },
\end{equation}
respectively, where $V(r)$ is the potential of the central force
\cite{prec}. The potential term consists of the Post-Newtonian
potential,
\begin{equation}
V_{PN}(r)=-\frac{\alpha }{r}-\frac{3\alpha ^{2}}{mr^{2}},
\end{equation}
plus the contribution from the general coupling between matter and
geometry. Thus, one has
\begin{equation}
\frac{d\vec{A}}{d\theta }=r^{2}\left[\frac{ 6\alpha
^{2}}{mr^{3}}+m\vec{a} _{E}(r)\right] \vec{e}_{\theta },
\end{equation}
where it is also assumed that $\mu \approx m$. The change in
direction $\Delta \phi $ of the perihelion for a variation of
$\theta $ of $2\pi $ is obtained as
\begin{equation}
\Delta
\phi =\frac{1}{\alpha e} \int_{0}^{2\pi }\left\vert \dot{\vec{L}}\times d%
\frac{\vec{A}}{d\theta }\right\vert d\theta ,
\end{equation}
 and is given by
\begin{eqnarray}\label{prec}
\Delta \phi& =&24\pi ^{3}\left( \frac{a}{T}\right) ^{2}\frac{1}{1-e^{2}}+\frac{%
L}{8\pi ^{3}me}\frac{\left( 1-e^{2}\right) ^{3/2}}{\left( a/T\right) ^{3}}\times \nonumber\\%
&&\int_{0}^{2\pi }\frac{a_{E}\left[ L^{2}\left( 1+e\cos \theta \right)
^{-1}/m\alpha \right] }{\left( 1+e\cos \theta \right) ^{2}}\cos \theta
d\theta ,\nonumber\\
\end{eqnarray}
where the relation $\alpha /L=2\pi \left( a/T\right) /\sqrt{%
1-e^{2}}$ is used. The first term of this equation corresponds to
the GR prediction for the precession of the perihelion of planets,
while the second gives the contribution to the perihelion
precession due to the presence of the new coupling between matter and
geometry.

As an example of the application  of Eq.~(\ref{prec}), one
considers the case for which the extra-force $a_E$ may be
considered constant --- an approximation that might be valid for
small regions of the space-time. Thus, through Eq.~(\ref{prec}),
one obtains the perihelion precession
\begin{equation}\label{prec1}
\Delta \phi =\frac{6\pi GM_{\odot}}{a\left( 1-e^{2}\right)
}+\frac{2\pi a^{2} \sqrt{1-e^{2}}}{GM_{\odot}}a_{E},
\end{equation}
resorting to Kepler's third law, $T^2=4\pi ^2a^3/GM_{\odot}$.

For Mercury, $a=57.91\times 10^{9}$ m and $e=0.205615$,
respectively, while $M_{\odot }=1.989\times 10^{30}$ kg: the first
term in Eq. (\ref{prec1}) gives the GR value for the precession
angle, $\left( \Delta \phi \right) _{GR}=42.962$ arcsec per
century, while the observed value is $\left(\Delta \phi
\right)_{obs}=43.11\pm0.21$ arcsec per century \cite{merc}.
Therefore, the difference $\left(\Delta \phi
\right)_{E}=\left(\Delta \phi \right)_{obs}-\left( \Delta \phi
\right) _{GR}=0.17$ arcsec per century can be attributed to other
physical effects. Hence, the observational constraints requires
that the value of the constant extra acceleration $a_E$ must
satisfy the condition
\begin{equation}
a_E\leq 1.28\times 10^{-11}\;{\rm m/s}^2.
\end{equation}
This value of $a_E$, obtained from the solar system observations,
is somewhat smaller than the value of the extra-acceleration $
a_{0}\approx 10^{-10}$ m/s$^{2}$, necessary to account for the
Pioneer anomaly \citep{Bertolami:2007gv}. However, it does not
rule out the possibility of the presence of some extra
gravitational effects acting at both solar system and galactic
scale, since the assumption of a constant extra-force may not be
correct on large astronomical scales.

\section{Conclusions and Outlook}\label{Sec:VII}

In this contribution we have discussed a wide range of implications of the
gravity model action, Eq.~(\ref{model}), whose main feature is the
non-minimal coupling between curvature and the Lagrangian density
of matter (or a function of it, in Section VI). This exhibits an
extra force with respect to the GR motion, as well as the
non-conservation of the matter energy-momentum tensor. The
prevalence of these features for different choices for the matter
Lagrangian density was discussed in Section III. In Section IV,
the specific features of the associated scalar-tensor theory were
discussed --- and it was shown that the model is consistent with
the observational values of the PPN parameters, namely $\be = \ga
= 1$, to zeroth-order in $\la$. In Section V, we consider the
impact of the novel coupling on the issue of stellar equilibrium.
It is shown that, for the simplest model of the Sun, the effect of
the new coupling on the central temperature is smaller than 1 \%,
which is consistent with the uncertainty of current estimates.
Finally, in Section VI, a general function of the matter
Lagrangian density has been introduced, and the value of the
resulting extra force obtained, $a_E \leq 10^{-11}~{\rm m/s}^2$.

Of course, further work is still required in order to quantify the
violation of the Equivalence Principle introduced by the model
under realistic physical conditions. A low bound for the coupling
$\la$, would justify the results discussed in this work, which are
first order in $\la$. Implications of the discussed model in what
concerns the issue of singularities are still to be addressed, as
well as the impact that the new coupling term might have on the
early Universe cosmology.

We would like to close this contribution with our best wishes to
our colleague Sergei Odintsov, on the occasion of his 50th
birthday.

\begin{acknowledgments}

O.B. acknowledges the partial support of the Funda\c{c}\~{a}o para
a Ci\^{e}ncia e a Tecnologia (FCT) project $POCI/FIS/56093/2004$.
The work of T.H. was supported by a GRF grant of the Government of
the Hong Kong SAR. F.S.N.L. was funded by FCT through the grant
$SFRH/BPD/26269/2006$. The work of J.P. is sponsored by FCT
through the grant $SFRH/BPD/23287/2005$.

\end{acknowledgments}

\end{document}